# Superconductivity and unusual magnetic behavior in amorphous carbon

Israel Felner

Racah Institute of Physics, The Hebrew University, Jerusalem, 91904, Israel

**Abstract**

Traces of superconductivity (SC) at elevated temperatures (up to 65 K) were observed by magnetic measurements in three different inhomogeneous sulfur doped amorphous carbon (a-C) systems: (a) in commercial and (b) synthesized powders and (c) in a-C thin films. (a) Studies performed on commercial (a-C) powder which contains 0.21% of sulfur, revealed traces of non-percolated superconducting phases below $T_c$ = 65 K. The SC volume fraction is enhanced by the sulfur doping. (b) a-C powder obtained by pyrolytic decomposition of sucrose did not show any sign for SC above 5 K. This powder was mixed with sulfur and synthesized at 400 ºC (a-CS). The inhomogeneous products obtained, show traces of SC phases at $T_C$= 17 and 42 K. (c) Non-superconducting composite a-C-W thin films were grown by electron-beam induced deposition. SC emerged at $T_c$ = 34.4 K only after heat treatment with sulfur. Other parts of the pyrolytic a-CS powder, show unusual magnetic features. (i) Pronounced irreversible peaks around 55-75 K appear in the first zero-field-cooled (ZFC) sweep only. Their origin is not known. (ii) Unexpectedly these peaks are totally suppressed in the second ZFC runs measured a few minutes later. (iii) Around the peak position the field-cooled (FC) curves cross the ZFC plots (ZFC>FC). These peculiar magnetic observations also ascribed to a-CS powder prepared from the commercial a-C powder and are connected to each other. All SC and magnetic phenomena observed are intrinsic properties of the sulfur doped a-C materials. It is proposed that the a-CS systems behave similarly to well known high $T_C$ curates and/or pnictides in which SC emerges from magnetic states.

PACS numbers: 74.10+v, 74.70.Wz, 75.50.Kj, 75.75.-c,

**I. Introduction.**

Interest in nano-sized particles has increased in the last few years by virtue of their potential for applications. These materials such as amorphous carbon (a-C), with dimensions intermediate between molecules and bulk solids are intriguing due to their novel electronic features which depend on the particle size, topology, surface conditions etc. Unlike other group IV amorphous semiconductors, such as amorphous-Si or amorphous-Ge, which have a tetrahedral $sp^3$-bonded random-network matrix, typical a-C powders and/or a-C films have a mixture of $sp^2$ and $sp^3$ bonding. Changing the $sp^2/sp^3$ ratio tunes the band gap of a-C between that of diamond (100% $sp^3$, band gap Eg=5.4 eV) and graphite (100% $sp^2$, semimetal with Eg=0). Amorphous carbon contains partially graphitized carbon fragments that possess both negative and positive curvatures; therefore, a-C is a promising candidate for searching for (i) superconductivity (SC) and (ii) for very unusual magnetic behavior. [1-2]

*Superconductivity.* (a) In recent studies, we have demonstrated that two samples extracted from an old commercial inhomogeneous a-C powder exhibit traces of non-percolative SC at $T_C$=32 and ~ 65 K [2-3]. This powder contains tiny amount of sulfur (around 0.2 at%) and it was suggested that SC is due to unknown a-C-sulfur (a-CS) phases. The existence of SC in this commercial a-C powder encouraged us to search for high $T_C$ SC in a synthesized a-C source. (b) Indeed, synthetic a-C material obtained by melting of pure sucrose ($C_{12}H_{22}O_{11}$) which was mixed with sulfur and heated to 400 °C, exhibited traces of two SC phases: at $T_C$= 17 K and 42 K.[4] (c) The third system we studied, comprise thin composite films of granular a-C and tungsten prepared by Electron-Beam Induced Deposition (EBID). After treatment with sulfur at 250 ºC the resultant aC-W-S films showed clear, yet still small, traces of SC at $T_c$ =34 and 40 K.[3,5] The above three systems, prove unambiguously the existence of SC in sulfur-doped a-C materials, although the exact compositions of the various SC phases obtained are as yet not known, These experiments are in good agreement with theoretical expectations, which suggest that both adsorbed sulfur and structural disorder can locally induce extra carriers (doping effect) into graphite/graphene and therefore trigger or enhance SC. [6]

*Unusual magnetism.* A growing number of carbon-based materials (CBM) have been found to exhibit ferromagnetic (FM) or a FM-like behavior, which may be attributed to the mixture of carbon atoms with $sp^2$ and $sp^3$ bonds as discussed above. [7-9] It is still unclear, which are the most crucial factors that stabilize the FM-like state. The experimental evidences accumulated so far indicate that structural disorder, topological defects produced by proton irradiation, [9] or alternatively graphitic fragments with positive or negative curvatures, [10] as well as adsorbed foreign atoms such as hydrogen [11] and/or oxygen [12] may be the FM origin for FM in CBM. Super-paramagnetic-like behavior has been observed in a-C obtained by chemical synthesis or by pyrolysis. [13-14]

The discovery of FM in carbon nano-foam, which possesses a low density and a large surface area, sheds some light on the FM-like state in CBM. [15-17] The zero-field-cooled (ZFC) and field-cooled via cooling (FCC) or via warming (FCW) magnetization curves display a complex behavior with a maximum in ZFC branch whose position and size are sample dependent. The ZFC and FC curves merge at room temperature indicating that the magnetic transition is at a higher temperature. Carbon nano-foam shows an enhanced magnetization immediately after the synthesis, most of which is actually lost on a very short time scale. The simulated structures show that trivalent carbon atoms are the major source for magnetism in carbon nano-foam.[18] Similar to carbon nano-foam, a-C powder possesses low density and high surface area. However, to the best of our knowledge, no systematic magnetic measurements have been performed on commercial and/or on chemical synthesized a-C powders.

Here, we review unusual magnetic properties of three a-C materials obtained from different sources: (a) commercial and (b) synthetic a-C powders and (c) a-C thin films obtained by EBID. [3-5] All investigated materials appear to be inhomogeneous. That means that the magnetic features measured on different samples taken from the same batch or source may differ each from other. The scattered results obtained on both powders can be divided into three groups: (i) in most samples measured, no any significant magnetic features are observed. An increase at low temperatures is obtained due to the paramagnetic (PM) impurities. The ZFC and FC magnetization curves depict the FM features of magnetite ($Fe_3O_4$) which is ferri-magnetically ordered up to 853 K. (ii) A few samples have shown traces of SC at various temperatures as discussed above. Due to its amorphous nature, we did not succeed to compact all nano-size powder particles into pellets, hence no resistivity measurements were possible. (iii) Other samples (from the same batch) exhibit pronounced peaks in their ZFC curves at 50-80 K. Heating both a-C powders with sulfur (a-CS) at 400 °C does not change significantly the sulfur content. However, unexpectedly in these a-CS samples two anomalies which are connected to each other are observed. First, around the peaks position, the ZFC branches are much higher than the FC curves, thus at a certain temperature range ZFC>FC. Second, this complex behavior is actually irreproducible and disappears in the second ZFC run. The observed phenomena can neither be ascribed to conventional FM nor to impurities (such as magnetite and/or oxygen), and it is believed that they are an intrinsic property of the a-CS materials.

In this review, we combine the already accumulate published data [3-5] together some with new unpublished results. We show here, the most prominent magnetic features obtained on representative a-C and a-CS samples which exhibit SC or peculiar magnetic features. This paper is divided into three parts. First, the magnetic features of commercial a-C powders are presented. Next, we describe the synthesized melted a-C powder (without sulfur) which serves as a blank material and then, the SC phases and the peculiar magnetic behavior obtained in a-CS materials. The third part deals with the observed SC state in fabricated a-CS thin films.

**II. Experimental Details.**

Two commercial a-C powders have been used. (1) a-C was manufactured by Fisher (C190-N) 78 years ago. Most of the experiments reported here have been performed on this powder. ICP as well as Mössbauer studies show the presence of about 300 ppm of magnetite ($Fe_3O_4$) and of other PM impurities. (2) Commercial a-C powder was manufactured by Prolabo (9009) as charbon R.P. The pyrolytic a-C was obtained by multi-stage thermal treatment of pure sucrose ($C_{12}H_{22}O_{11}$) as described in detail in Ref. 19. First, (186 °C, 3 hours) the volatile components and water were removed from the viscous melt (caramel). Solid powder was obtained after further heating the melt at 300, 400 and 650 °C for 1 hour (each). Mixtures of a-C powders and sulfur (Aldrich Chemical Company, Inc.) or boron powders in a weight ratio of 2:1 (for commercial) and 10:1 or 10:3 (for pyrolytic) were well mixed, pressed and heated at 250 or 400 °C for 20-24 hours in evacuated quartz tube. The granular aC-W thin films were grown by employing the EBID technique, as described in detail in Refs. 5 and 20. $W(CO)_6$ precursor molecules were introduced into the chamber of a scanning electron microscopy (SEM, e-line, Raith) and using 10 keV electron-beam energy, resulting in a weakly insulating ($dR/dT < 0$) granular aC-W film. The aC-WS composites were obtained by heating these films with sulfur in evacuated quartz tube at 250 °C. The materials, were structurally and chemically characterized by x-ray diffraction (XRD), SEM, energy dispersive spectroscopy (EDS) JOEL JSM-7700 SEM. Trace elements analysis was performed by inductively coupled plasma (ICP) mass spectrometer (Perkin-Elmer ICP-OES model 3300) of acid extracts. Magnetization measurements on samples mounted in gel-caps at various applied magnetic fields (H) in the temperature interval 5 K < T < 300 K, have been performed using a Quantum Design superconducting quantum interference device (SQUID) magnetometer. The differential SQUID sensitivity is $10^{-7}$ emu. Prior to recording each ZFC curve, the magnetometer was adjusted to be in a *"real"* H = 0 state.

## III. Experimental results.

### *(i). Pristine commercial (Fisher) amorphous carbon (a-C).*
*Structural and chemical characterization.* This commercial a-C powder was examined by XRD studies and no distinct sharp crystallographic reflections were observed, indicating its amorphous state. EDS chemical analysis shows: oxygen 2.44(1), sodium 0.30(1) and sulfur 0.21(1) at%) as extra elements (Fig. 1). Their distribution among the measured particles is not uniform. No other elements have been observed. SEM images (Fig. 2) show no systematic appearance of any crystallized particles, confirming their amorphous state. The deduced average particles size is 9-10 nm.

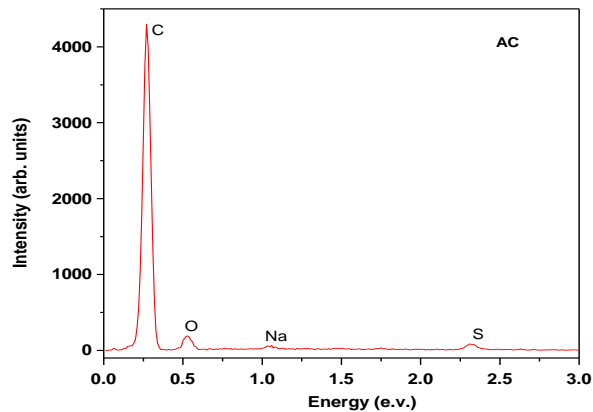

Fig. 1. EDS spectrum of commercial a-C sample. Note the presence of a small amount of sulfur.

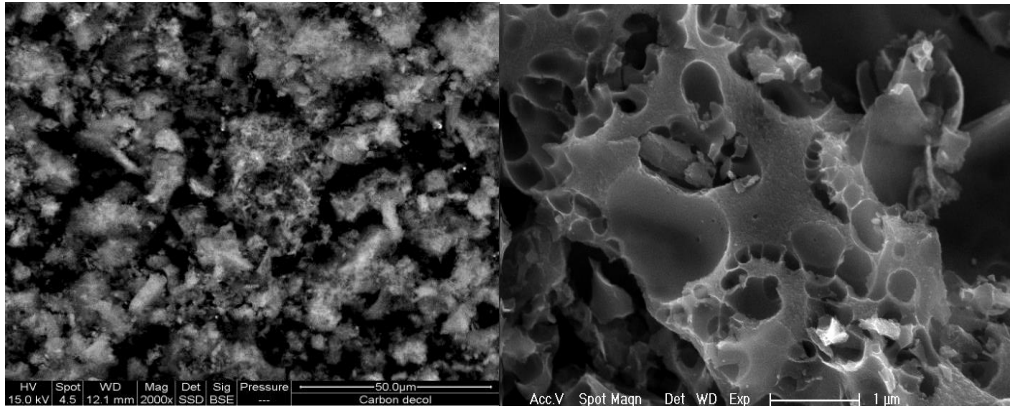

Fig. 2. Scanning electron microscopy (SEM) images of pristine commercial a-C powder.

ICP studies for trace element analysis were carried out on this a-C powder which was dissolved concentrated $HNO_3$ and distilled water (10 cc each) in an alumina crucible. A blank crucible (without the material) was treated simultaneously under the same conditions. The sample was scanned several times. The trace elements obtained are divided into non-magnetic and magnetic elements. The concentrations in ppm of the non-magnetic elements are: V=2.0, Zn=7.1, Cu=11.1, Al=212.7 and Na=4625 (note, the similarity to the EDS value, Fig. 1). The magnetic element concentrations are: Ni=2.8, Mn=133 and Fe=360 ppm, (the total amount is 496 ppm). Long-term RT $^{57}$Fe Mössbauer spectroscopy study shows two magnetic sub-spectra (not shown) which are attributed by a least square fit analysis to ~300 ppm of magnetite. This value fits well with the ICP value for Fe.

*Magnetic studies.* Comprehensive temperature dependence of magnetic moments on twenty three a-C samples under various applied fields, have been performed, by taking 13-20 mg samples directly from their glass container in which they were stored for more than seven decades. The scattered results obtained are divided into three groups: (i) in most samples measured, the ZFC and FC magnetization curves depicted the magnetic features of $Fe_3O_4$ and the PM impurities. (ii) Three samples have shown traces of SC at $T_C$= 32 K [2] and at 65 K. (iii) Four samples show pronounced peaks at 65-85 K in their ZFC branches when measured at low H only. For the sake of brevity, we present here the most prominent magnetic features obtained on representative samples of these three groups. The a-C "Prolabo" powder did not show any SC traces and basically exhibited typical magnetic features observed in groups (i) and (iii).

(i) Fig. 3 shows two ZFC(T) and one FC(T) curves for a-C (sample 13) measured at H= 50 Oe. The protocol of the measurements is as follows. (I) The sample was cooled to 5 K at H=0, then H was applied and first ZFC plot via heating was recorded. (II) The sample was cooled down to 5 K under H, to trace the FC curve up to 180 K. (III) H was switched off and the second ZFC plot was tracked. Fig. 3 shows that the two ZFC curves are identical indicating a reproducible ZFC process. That is in sharp contrast to the peculiar magnetic behavior of the heat treated materials describe below. The increase at low temperatures is due to PM impurities, whereas the bifurcation of curves at 180 K is caused by $Fe_3O_4$.

(ii) ZFC(T) and FC(T) curves measured at 40 Oe on a-C (sample 16) are depicted in Fig. 4. The irreversibility observed at $T_C$= 65±1 K is attributed to the onset of SC. The shielding fraction (SF) of ~ 0.013% is estimated from the almost constant part of the ZFC plot. This value fits well with the low SF obtained for the SC phase (sample 1) at $T_C$= 32±1 K [2]. It should be noted, however, that since no corrections for carbon magnetism and for $Fe_3O_4$ were made, this SF provides only a lower bound for the SC volume fraction. Nevertheless, the small SF value obtained suggests a localized SC state in this a-C sample.

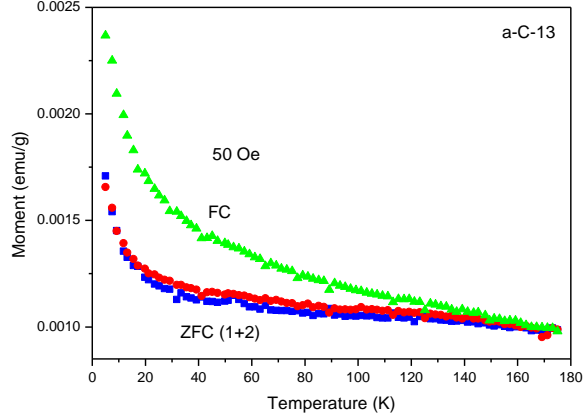

Fig. 3 ZFC and FC plots of commercial a-C (sample 13) measured at 50 Oe. Note the similarity of two ZFC curves.

The very clear drop at ~ $T_c$ in the FC curve is associated with the magnetic-flux expulsion (known as the Meissner effect) which is the fingerprint and an unambiguous experimental evidence for SC. Similar SC features were observed in another a-C powder (sample No. 18) with $T_C$= 66±1 K. These two $T_C$ values (at 32 and 65 K) suggest that the a-C powder contains (at least) two SC phases, a phenomenon which is well expected for inhomogeneous systems. Due to its amorphous nature, we did not succeed to compact the nano-size powder particles into a pellet, hence no resistivity measurements were possible.

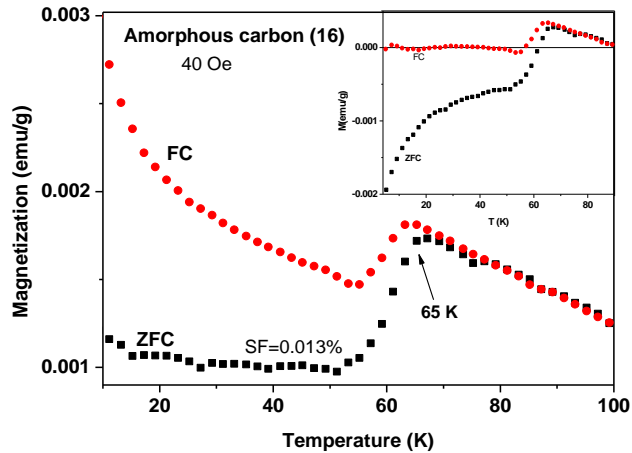

**Fig. 4**: ZFC and FC temperature-dependent magnetization plots of a-C (sample 16) measured at 40 Oe. The inset shows the M(T) curves obtained after subtraction of the paramagnetic contribution. A clear SC transition is observed at 65 K.

At low temperatures, both ZFC and FC branches (Fig. 4) exhibit a typical PM shape and adhere closely to the Curie-Weiss (CW) law: $\chi(T) = \chi_0 + C/(T-\theta)$, where $\chi_0$ is the temperature independent part of the susceptibility $\chi$, $C$ is the Curie constant, and $\theta$ is the CW temperature. The PM parameters extracted for both branches are: $\chi_0$ =9.1×10$^{-4}$ and 1.1·10$^{-3}$ emu/g Oe, C = 3.23×10$^{-2}$ and 2.28×10$^{-2}$ emu K/g Oe and $\theta$ = - 0.5 and 3.2 K respectively. Fig. 4 (inset) shows the two ZFC and FC curves obtained after subtracting the PM contribution of each branch. The low temperature flat curve of the FC curve is readily observed. Due to the PM background, considering the low SF and the presence of $Fe_3O_4$, further characteristic parameters of the SC state were not evaluated.

Since the a-C powder contains ~ 0.21 at% of S and bearing in mind that adsorbed sulfur can trigger and/or enhance SC in graphite [21-24], a mixture of this a-C powder with sulfur (ratio 2:1) was heated at 250°C for 24 hours in an evacuated quartz tube. The S concentration in this sample (a-CS) increased to ~10.3 at%, whereas concentrations of all other extra elements (Fig. 1) practically remain the same. The SC nature of this material was extensively presented in Ref. 2 and for the sake of comparison, its low temperatures range data are presented in Figs. 5-6. Since the pristine a-C material is not homogeneous, it was expected that this synthesized sample, would produce more homogeneous product. It appears, however, that the product remains inhomogeneous, which means that parts taken from the same synthesized batch, exhibit different magnetic behavior as discussed later.

In Fig. 5 the ZFC, FCC and the remnant magnetization ($M_{rem}$) of this a-CS sample are shown. $M_{rem}$ which is related to the trapped magnetic flux vortices, was recorded after the FCC to 5 K, when the field was switched off. The irrelevant temperature-independent background magnetization of $Fe_3O_4$ (0.037 emu/g) was subtracted from all three branches. Both ZFC and $M_{rem}$ show a pronounced step-like feature and provide strong evidence for a type II SC state below $T_c = 38 \pm 1$ K. Importantly, the Meissner effect, though smaller in magnitude is well observed below $T_c$. As expected for SC, where the diamagnetism originates from screening super-currents, the ZFC(T) moments are negative, and the estimated SF deduced 0.15% , which is an order of magnitude higher than that of a-C described above. Since the a-C and a-CS samples differ only in their sulfur contents, it is reasonable to assume that the enhanced SF is triggered by sulfur.

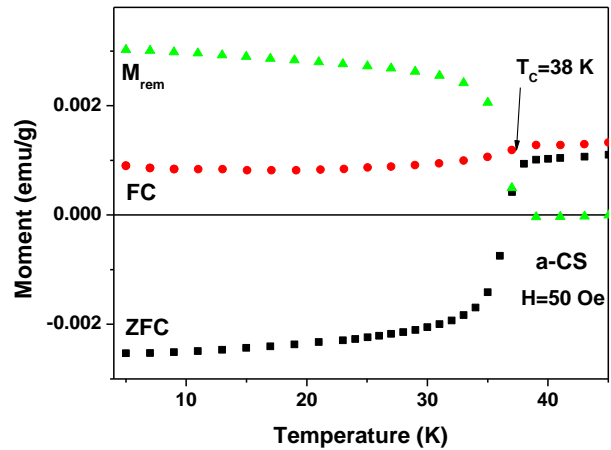

.
Fig. 5. ZFC, FCC (via cooling) and the remnant magnetization a-CS measured at 50 Oe. The background magnetization due to $Fe_3O_4$ is subtracted from all three branches.

Fig. 6 shows the hysteresis loop measured at 5 K up to H = 50 kOe. The PM behavior is evident from the reversible non-saturated high field portion of curve, giving the PM susceptibility $\chi \approx 1 \cdot 10^{-5}$ emu/g Oe, which is similar to that measured for the a-C sample. The remnant magnetization value $M_0$~ 0.24 emu/g, is much higher than $M_{rem}$, (Fig. 5) due to higher flux vortices trapped at the upper critical field ($H_{c2}$). At low fields Fig. 6 (inset), the virgin M(H) curve decreases linearly with H and deviates from linearity at the lower critical field: $H_{c1}$ ~ 250 Oe. The detailed $H_{c1}(T)$ as well as $H_{c2}(T)$ phase diagrams were presented in Ref. 2.

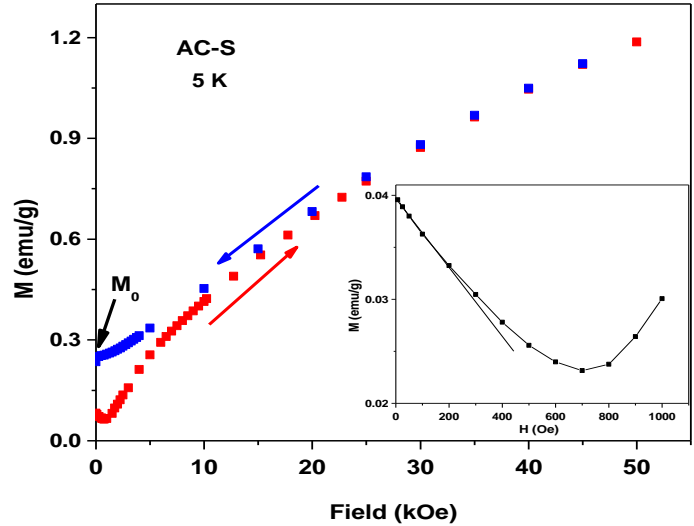

Fig.6. M(H) hysteresis loop of a-CS measured at 5K. The inset expands the low field values

(iii) Four a-C samples exhibit a pronounced peak in their ZFC branches only. Fig. 7 shows typical ZFC(T) and FCC(T) curves for sample No. 5 measured at 50 Oe, in which the peak at 67 K, can easily be seen. The FCC(T) branch is very similar to that observed in Fig. 3. Here again, the bifurcation of the two curves (related to $Fe_3O_4$) is observable at low applied field (50-100 Oe). At higher fields (0.5-1 kOe) the peak is also observed in the FCC(T) curve where both branches coincide into one plot (Fig. 7 inset).

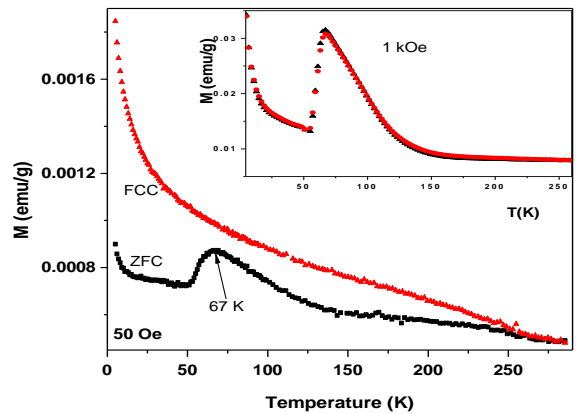

Fig. 7. ZFC and FCC of a pristine a-C sample (sample 5) measured at 50 Oe and at 1 kOe (in the inset).

The general trend of the peak position field dependent is depicted in Fig. 8. For sample 18, the peak at 79 K under 50 Oe (not shown), shifts with H to higher temperatures. The second commercial a-C "Prolabo" powder, did not show SC traces but basically exhibited the magnetic features obtained for groups (i) and (iii).

Peaks around T~ 50-55 K may originate from the adsorbed oxygen.[25] Actually, traces of solidified oxygen are visible and appear in both ZFC and FC branches (see Figs. 11,13). However, they are negligible in comparison with the main magnetization peak. In addition, the peaks in Figs. 7-8 cannot be attributed to oxygen since they appear at much higher temperatures and their amplitude is bigger by four orders of magnitude. It is thus assumed that peaks observed are intrinsic properties of the inhomogeneous a-C powder.

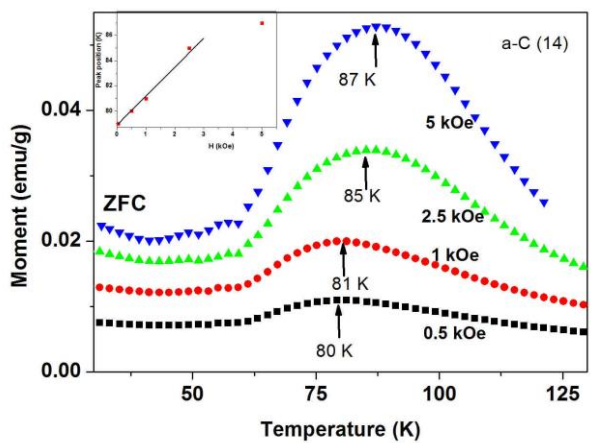

Fig. 8. ZFC plots of another a-C material (sample no. 14) at various applied fields. Note the shift of the peak position with the field.

Peaks in the magnetization curves appear in carbon nano-foam [16-17] and in pure bulk graphite.[26] In carbon nano-foam the peak position, is very sensitive to preparation conditions, and the general magnetic behavior is consistent with a spin-glass-type ground state, thus the peaks indicate the spin freezing temperatures.[17] In graphite, the system is highly irreversible and the peaks at ~115 K in both ZFC(T) and FC(T) branches are attributed to weak coupling between FM regions related to inhomogeneous defects and the FM inside the defect regions.[26] None of these systems, resembles the behavior of a-C examples shown here (Figs. 7-8).

*(ii). Fabricated pyrolytic amorphous carbon.*
*Structural and chemical haracterization of a-CS.* This pyrolytic a-C powder was obtained by multi-stage thermal treatment of pure sucrose. The initial Fe content was < 3 ppm. Mixtures of this a-C and sulfur powders (weight ratio of 10:1 or 10:3) were pressed and heated at 400 °C for 24 hours as described above. XRD pattern taken on a-CS powder does not show any distinct sharp crystallographic reflections, indicating its amorphous nature (Fig. 9). EDS chemical analysis of a-CS shows that the average concentrations of sulfur and oxygen are 13% and is 12%. No other elements have been observed. ICP measurement on a-CS, reveals that the trace elements are: Zn=2.7, Al=3.2, Ni=4.3 ppm. The Fe concentration is increased to 28 ppm presumably appearing as $Fe_3O_4$. Thus the main significant extra elements in a-CS are: oxygen and Fe. SEM images taken from various a-CS samples show no systematic appearance of any crystallized particles, confirming their amorphous state.[4]

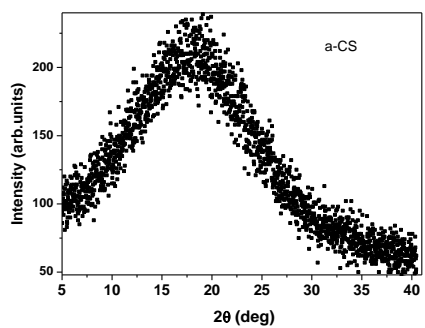

Fig. 9. XRD pattern of pyrolytic a-C synthesized with S at 400 °C (a-CS).

*Magnetic studies of pyrolytic a-C material (undoped).* The ZFC and FC branches of the pyrolytic un-doped a-C powder, which serves as a blank material, are depicted in Fig. 10. Generally speaking, this a-C powder is diamagnetic. Due to the presence of a tiny amount of unpaired localized spins, the low temperatures part of ZFC branch follows the CW law and the PM parameters extracted are: $\chi_0 = -4.4 \cdot 10^{-7}$ emu/g Oe, $C = 8.32 \cdot 10^{-6}$ emu K/g Oe and $\theta = 0.16$ K. Note that the positive moments in Fig. 10 are two orders of magnitude lower that the value obtained for commercial a-C shown in Fig. 3. The bifurcation of the two branches at 120 K, the so called-Verwey transition of $Fe_3O_4$ is barely observed in the present studies.

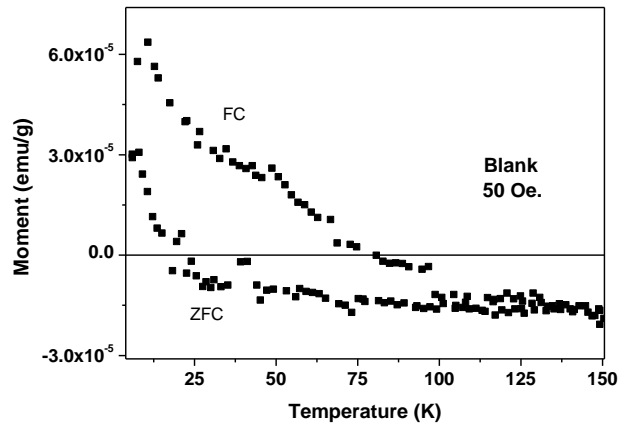

Fig. 10. Temperature dependence of the ZFC and FC branches of the un-doped pyrolytic a-C.

*Magnetic properties of a-CS (10:1 sulfur doped) samples*

Comprehensive magnetic measurements have been performed on several a-CS powders (fabricated from sucrose), by taking 13-20 mg samples from their container in which they were stored for 1-2 years. Here again, the a-CS materials are not homogeneous and exhibits scattered results. Two samples show a non-trivial SC-like behavior at $T_C = 17$ K and 42 K. Few other samples are magnetic and exhibit pronounced peaks in their *first* ZFC curves in the range of 50-60 K and the rest did not show any significant magnetic behavior.

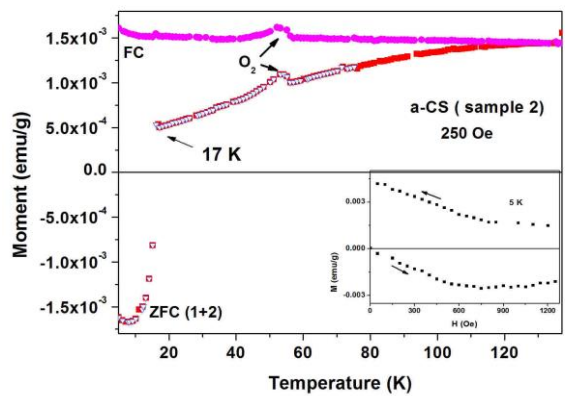

Fig. 11. ZFC(1) (bold, red), ZFC(2) (open, blue) and FC magnetization plots of a-CS (sample 2) measured at 250 Oe. The peaks in both branches are due to solidification of oxygen. The inset shows the typical SC hysteresis loop obtained at 5 K.

*Superconductivity.* (i) The ZFC(T) and FC(T) magnetization curves measured at 250 Oe on a-CS powder (sample 2) are depicted in Fig. 11. Before measuring the FC branch, the ZFC process was repeated and the second curve ZFC(2) obtained; it perfectly coincides with the first ZFC curve. The sharp drop of the two ZFC curves is attributed to the onset of SC at $T_C$=17 K. Fig. 11 also shows the signals correspond to oxygen in **both** ZFC and FC branches. The extracted small SF~ 0.016 % fits well with that deduced for the commercial a-C powder (Fig. 4). This value provides only the lower bound for the SF, but definitely suggests a localized SC state in this a-CS powder. The low M(H) regime measured at 5 K, exhibits the typical SC hysteresis loop (Fig. 11 inset). The measured remnant (Fig. 12) is an additional supporting evidence for the SC state.

(ii) In order the evaluate the sulfur effect on the SC state, another a-CS powder was heated at 250 °C for 2 hours and further at 380 °C for 3 more hours (assigned as a-CS(h)). EDS analysis shows that the sulfur content of a-CS(h) was reduced to 2.1%, whereas the oxygen concentration (13.1 %) remained basically unchanged. The ZFC and FC curves for this material are shown in Fig. 13. Here again, the sharp rise at 42 K is attributed to a SC phase. In contrast to Fig. 11, below $T_C$ the ZFC signal is positive and both ZFC and FC curves exhibit a typical PM shape. The values extracted by applying the CW law, are quite similar (within experimental errors) to values deduced for the blank material (Fig. 10). That means that the amount of PM impurity level as well as the oxygen concentration did not change during the heating processes. Due to the paramagnetic background and to the presence of $Fe_3O_4$, more characteristic SC parameters of the SC state were not evaluated. The magnetic peaks in both branches corresponding to oxygen are readily observed.

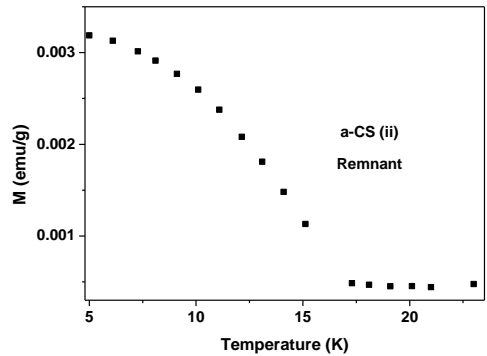

Fig. 12. Remnant magnetization of a-CS (sample 2)

In addition to the SC phases observed in *commercial* a-C and a-CS (Figs 4-6), we may propose (with high confidence) that the magnetic signals in Figs. 11-13 are very likely due to the presence SC phases in the *pyrolytic* a-CS powders; their exact composition is not yet known. This assumption gains support from our experiment on aC-W composite thin films, in which SC was induced only after sulfur was added (see below). For commercial a-CS, it was proposed that the adsorbed sulfur content *enhances* the SC state, [2] but Fig. 13 proves that $T_C$ of a-CS(h) is higher than that of a-CS (Fig. 5), although its sulfur concentration is lower.

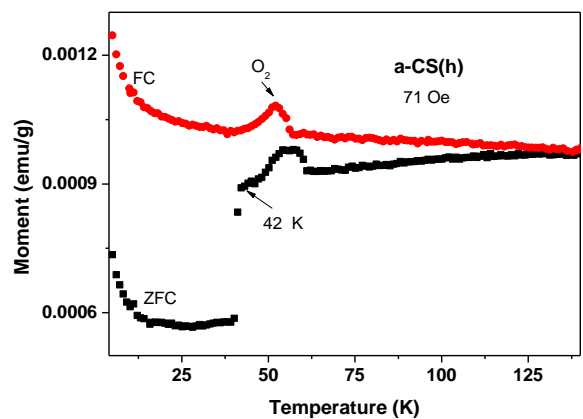

Fig. 13. ZFC and FC magnetization plots of a-CS(h) measured at 71 Oe. The peaks in both branches are due to solidification of oxygen.

*Peculiar Magnetic behavior of pyrolytic a-CS materials.* In what follows, we describe the magnetic properties of two inhomogeneous a-CS materials (differing in their initial sulfur concentrations) synthesized separately at 400 °C under the same conditions.

Two ZFC and FC processes measured at 500 Oe for a-CS material (a-C:S 10:) are presented in Fig. 14. In order to eliminate the PM contribution the lowest measured temperature is 20 K. The protocol in Fig. 14 is as follows. (a) The sample was cooled at H=0 to 20 K. Then, H was turned on to trace ZFC(i) branch up to 150 K. (b) Then, the FCC (I) process was applied down to 20 K (c) H was turned off, the sample was heated to room temperature and immediately cooled back to 20 K. Then, H was switched on to sweep the ZFC(ii) up to 70 K. (d) The FCC(ii) processes down to 20 K was completed.

Fig. 14 shows that the pronounced peak in ZFC(i) around 50 K has completely disappeared in the FC(i) plot. In contrast to Fig. 7, unexpectedly the FC(i) curve crosses the ZFC(i) one, at 42 and 53 K. Thus in this temperature range, ZFC(i) > FC(i) (**puzzle 1**). Moreover, this peak does not show up in the ZFC(ii) curve which was measured a few hours after ZFC(i) (**puzzle 2**). On the other hand, the two FC branches coincide with each other, indicating that the sample was not damaged or alternatively no residual magnetization was gained, in this short time period. The appearance of the peak in the ZFC(i) run only, is a rare phenomenon seldom observed.

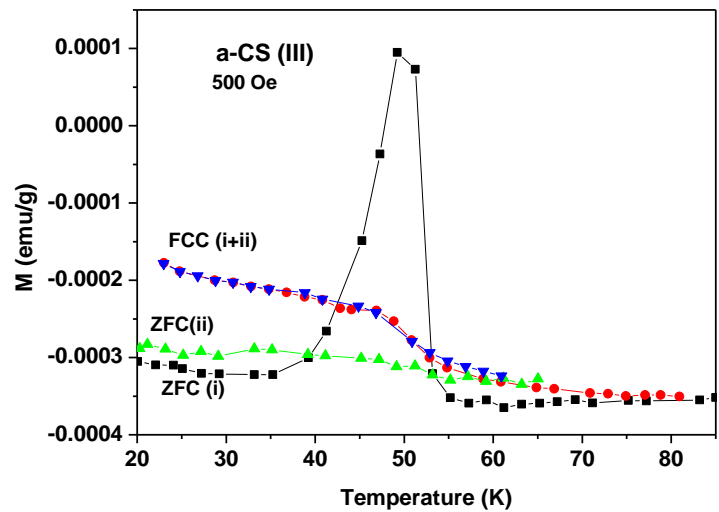

Fig. 14. (color online) Two ZFC and FC plots of a-CS (a-C:S is 10:1, sample 3) measured at 500 Oe. The ZFC(i) (in black) and FCC(i) (in red) curves were measured up to 120 K.

One may argue that this peak is related to adsorbed oxygen as shown in Figs. 11 and 13. This assumption is ruled out (as discussed earlier) for the following reasons. (*1*) The normalized peak magnitude is much higher than that measured in Fig. 13. (*2*) The peak related to oxygen always appears in **both** ZFC and FC branches (see Figs. 11 and 13). Since in the ZFC(i) process the sample was heated up to 120 K only, the absence of the peak in ZFC(ii) and/or in FCC(i) definitely excludes this assumption. By the same token, if this peak (in any way) is attributed to $Fe_3O_4$, it should also be observed in the FC branches as well. (*3*) One may argue that this peak is an artifact which arises from unknown experimental failures of the SQUID magnetometer or from the possibility that the sample was not measured under optimal conditions. To exclude this speculation, we have shown (Fig. 11) overlap of the two ZFC sequences in the SC a-CS material.

The peak in Fig. 14 is obtained by (1) heating the sample under an applied field. The question arises is whether this peak is caused by the heating process only or it is induced by H. In order to answer this question, we have performed similar measurements on another a-CS sample (sample 5) with higher initial sulfur content (10:3) shown in Fig. 15. (i) First, the sample was ZFC to 5 K and heated up to RT at H=0 and cooled back to 5 K. (ii) H was switched on to 50 Oe. In order to shorten the elapsed time, the ZFC(i) branch was measured up to 75 K only. (iii) the field was switched off and the sample was cooled back to 10 K to trace ZFC(ii) up to 160 K. (iv) The FCC process was measured down to 5 K. The bifurcation of ZFC(T) and FCC(T) around 120 K is caused by $Fe_3O_4$. Here again, the huge peak around 50 K, caused by the applied field, appears in the ZFC(i) branch only and the FC curve crosses the ZFC(i) branch. This proves that the two unusual magnetic phenomena, namely the phenomenon that ZFC>FC (**puzzle 1**) at the peak position and the peak disappearance in ZFC(ii) (**puzzle 2**), are reproducible regardless of the sulfur concentration and/or the external H.

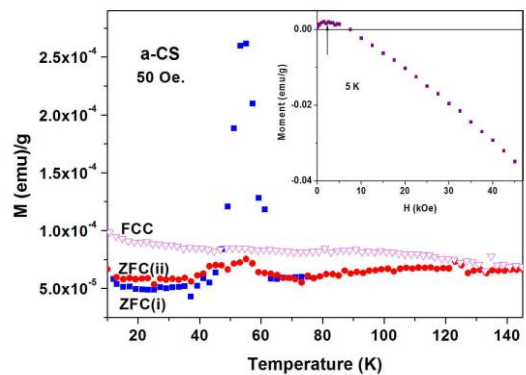

Fig. 15. Two ZFC and FCC branches of a-CS (a-C:S is 10:3) measured at 50 Oe. ZFC(i) (in blue) was measured up to 75 K. The ZFC(ii) plot was measured a few minutes after the ZFC(i) one. The inset shows the isothermal the M(H) plot measured at 5 K.

The isothermal M(H) data measured at 5 K of this sample are depicted in Fig15 (inset). Basically, this plot presents the diamagnetic character of the pyrolytic a-C material. However at low applied fields, due to the presence tiny amounts of magnetite and PM extra phases, the positive magnetization first increases and saturates at $M_S$ =0.0021 emu/g, then decreases and becomes negative. This $M_S$ value corresponds to 20.3 ppm of $Fe_3O_4$ ($M_S$ = 94.5 emu/g) and is in fair agreement with 28 ppm of Fe detected by ICP measurement. Similar M(H) plots were obtained in all other pyrolytic a-CS samples.

**Magnetic studies of commercial a-CS and a-CB samples.**
In order to confirm these two observed puzzles, similar measurements have been performed on: (1) commercial (Fisher) a-CS sample and (2) on commercial a-C powder mixed with boron powder (a-CB), all synthesized at 400 °C.

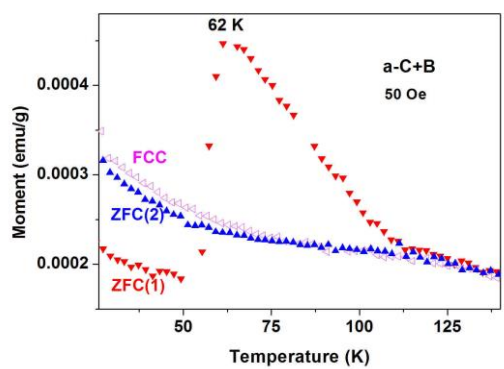

Fig. 16 Two ZFC and one FCC branches measured at 50 Oe for commercial a-C powder synthesized with boron powder.

Fig. 16 shows two ZFC and one FC curves on a-CB powder (instead of sulfur). The protocol of this study is similar to that described above. It is readily observed that the peak at 62 K measured in the first ZFC run disappears in the second one and that the FC crosses the ZFC (1). Measurements performed on commercial a-CS powders show similar behavior to that observed in Figs. 14-16. For the sake of brevity these results are not shown here.

As a final point of interest, we show in Fig. 17 three ZFC sequences performed on commercial a-CS powder. Here, in addition FC branch (not shown) measured right after the ZFC(1) one, a second ZFC process was measured two days. Here again, the pronounced peak at 78 K which appears in the ZFC(1) run is totally washed out in the second ZFC(2) run. The ZFC(2) plot totally coincides with the ZFC(1) curve, except for the temperature interval of the peak. Moreover, this material was kept in its capsule and re-measured 18 months later. Although its lower amplitude, the re-appearance of the peak in the ZFC(3) curve at 65 K is clearly shown.

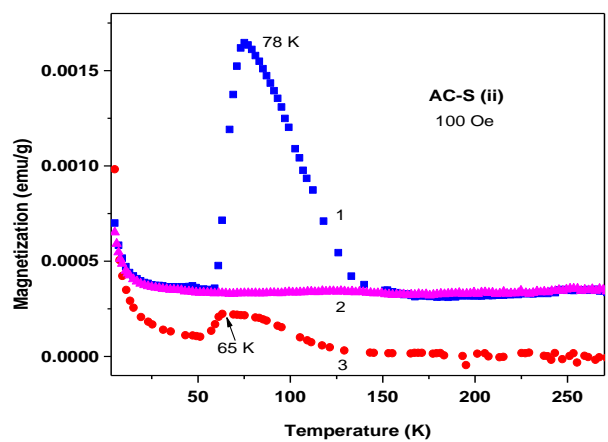

Fig. 17. Three ZFC curves for the same commercial a-CS measured at 100 Oe. The ZFC curves assigned as 2 and 3 were measured two days and 18 months after the first ZFC run.

The accumulated experimental results presented above are summarized as follows:
(1) All a-C and both a-CS materials are magnetically inhomogeneous.
(2) The magnetic features of a-C and a-CS regardless of whether the a-C is commercial or a pyrolytic product of sucrose contain also, PM impurities which are more pronounced at low temperatures and at higher external fields and a FM impurity phase most probable as $Fe_3O_4$.
(3) Some a-C samples exhibit an unconventional magnetic behavior. A pronounced peak appears in their ZFC curves. At high fields, the dominant PM contribution masks this peak. The peak position and its relative amplitude are sample dependent.
(4) In a-C, the external field shifts the peak position to higher temperatures.
(5) In a-CS, the ZFC(1) curves cross the FC ones and at certain temperatures ZFC > FC (puzzle 1). This observation is unique and to the best of our knowledge was not observed in the past. This anomaly appears at low applied fields only (Fig. 7 inset).
(6) The peaks observed in the first ZFC sweep only, disappears in the second ZFC process performed very shortly after the first run (puzzle 2).
(7) The peak is recovered after a long time interval.
(8) These salient features are also observed in boron doped materials.

This complexity does not permit measuring twice the same material or to compare results between two aliquots although taken from the same batch. The speculations that experimental failures and/or adsorbed oxygen and Fe impurities caused these peculiar magnetic phenomena have been ruled out. It is quite obvious that the two mentioned puzzles are inter-related to each other. Thus, puzzle 2 is a direct consequence of puzzle 1. Since the elusive peaks are always washed out after the first ZFC run, the FC branches, measured right after that, behave "normally" (Fig. 3). Due to impurity phases, they always lie above the second ZFC(II) branches as shown in Fig. 14. The elusive peaks origin is discussed later.

### *(iii) Superconductivity in sulfur doped a-C thin films.*

The above a-C powders prove unambiguously the existence of SC phases, caused by the presence of sulfur. The motivation for growing granular a-C-WS films was the hypothesis that tungsten inclusions, which are SC (with $T_c$ values that may vary significantly, see below), may increase the phase coherence between the localized SC islands in the a-C matrix. This effect is anticipated to raise the $T_c$ and the SF, akin to the $T_c$ enhancement effect found in bilayers of under-doped and over-doped $La_{2-x}Sr_xCuO_4$ high $T_c$ cuprate superconductors [27]. For this purpose, six a-C-WS thin films were grown by the EBID method, and two of them exhibit SC traces. The SF obtained is very sensitive to the growing procedure of the film.

Fig. 18 presents the magnetization curves measured at 89 Oe on one a-C-WS film and for comparison, its corresponding blank (a-C-W) film before the sulfur treatment. The blank film shows positive temperature-independent magnetization, without any sign of SC. The plotted ZFC and FC branches for the a-C-WS film were obtained after subtracting the blank magnetization from the measured data. The onset of SC at $T_c$ =34.4 K is readily observed. As expected below $T_c$, both the ZFC and FC magnetization curves are diamagnetic.

For higher applied fields (up to $H$ = 2 kOe) $T_c$ shifts to lower temperatures with a rate of $dH/dt$ = 0.45 kOe/K (not shown). Due to the field dependence of the positive blank magnetization, determination of the onset ($H_{C2}$) at higher applied fields was not accurate. Since neither the weight nor the composition of the SC phase is known, any estimation of the SF from the ZFC branch is misleading. Nevertheless, the relative small magnetic signals of both ZFC and FC, as well as the smeared ZFC curve, suggest an inhomogeneous superconducting state in this aC-WS film. Since the blank and a-C-WS film differ only in their sulfur contents, this a-C-WS film, serves as a second system in which the observed SC state is triggered by the sulfur presence. Due to its non-percolative small SC fraction, neither resistivity nor STM measurements have been performed.

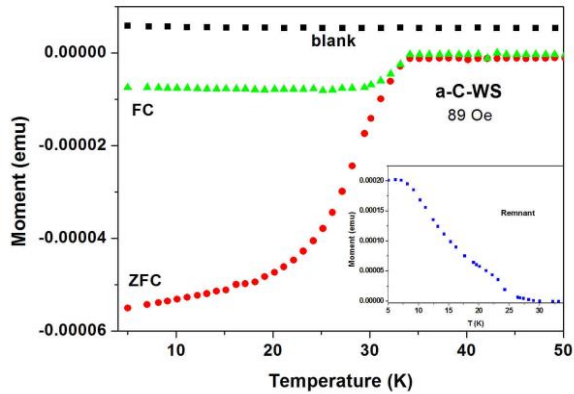

Fig. 18. ZFC and FC curves measured at 89 Oe on a-C-WS film and magnetization of the corresponding blank (aC-W) film. The inset shows the remnant magnetization of aC-WS.

The two other indications for SC in the a-C-WS films are: (i) The remnant magnetization that becomes zero very close to $T_c$, (Fig. 18 (inset)). This remnant was recorded after cooling the aC-WS film under 5 kOe from 38 K to 5 K and then H was switched off. It is reasonable to relate this remnant to the trapped magnetic flux (vortices) in the film. (ii) The two typical hysteresis loops measured at 5 and 10 K (Fig. 19). At low fields the *M(H)* plots deviate from linearity ($H_{c1}$) at ~ 600 and ~400 Oe for *T* = 5 and 10 K, respectively.

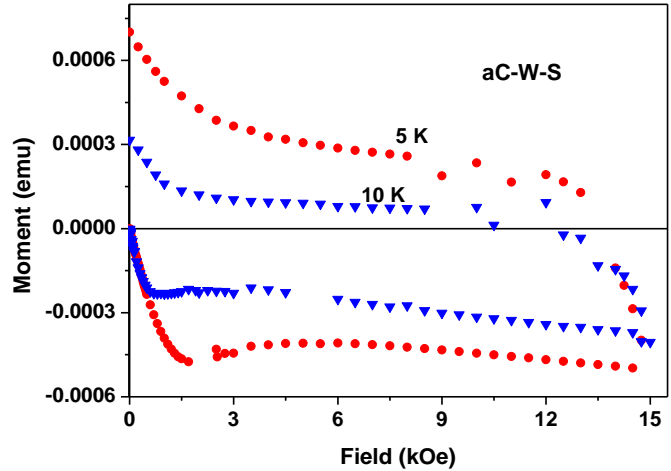

Fig. 19. Hysteresis loops of the aC-WS film measured at 5 and 10 K.

Figures 18-19, prove definitely the existence of a localized SC phase with $T_c$ = 34.4 K in the aC-WS film. The SC features are significantly clearer and more conclusive. In particular, the SC transition is sharp and even more significantly is that the Meissner effect in the FC branch is clearly observed. This may possibly stem from the existence of W and to the absent of PM and/or FM impurity phases. The W islands probably do not contribute by themselves to the onset of SC in the film, since SC appears only after the sulfur treatment. In addition, the maximal reports critical temperature of any composite comprising W and a non-SC material doesn't exceed 6.5 K. The $T_c$ of bulk α-phase W is ~15 mK, whereas in the less stable β-phase the reported $T_c$ values are in the range 1-4 K [28]. Significant enhancement of $T_c$, up to ~ 6.5 K, was found in films deposited using focused ion-beam (EBID), also using W(CO)$_6$ [18], whereas in amorphous W-Si multilayers the maximal achieved $T_c$ is 4.2 K [29]. These $T_c$s are significantly lower than 34.4 K found here. The role of the W inclusions is probably in enhancing the phase coherences by Josephson coupling nearby localized SC regions. As is well known, SC involves two imperatives the binding of electrons into pairs, which in our case may take place in the aC-S regions, and the establishment of phase coherence between them, presumably mediated more effectively by the W inclusions.

**IV Discussion**

*Superconductivity*. The above results provide clear experimental evidence for the existence of a tiny type II SC phase in sulfur doped a-C, at $T_C$ ranging from 17 to 65 K. SC was observed in **three** different a-C based systems all doped with sulfur, indicating that the SC state stems from a-CS phases. Both commercial and fabricated a-C materials, exhibit an inhomogeneous nature, namely, parts taken from the same batch, exhibit different magnetic behavior. (i) The pristine commercial a-C powder contains a small amount of S (0.21 at %), and the pronounced drop in both ZFC and FC branches (Fig. 4) can be attributed to traces of a SC phase below $T_c$=65 K. The deduced SF is 0.013% only. By heating a mixture of another a-C powder with sulfur at 250 °C, the sulfur concentration increases (10.3 at %) and as a result the SF increases by an order of magnitude to 0.15%, although $T_c$ was lowered to 38 K. This relative higher SF enabled the extraction of the upper and lower critical fields phase diagrams, both prove unambiguously the existence of a type II SC phase in the a-CS materials .[2] (ii) The second a-C source was obtained by pyrolysis of sucrose. This powder is basically diamagnetic. However, when synthesized with sulfur at 400 °C, traces of a SC phase at $T_C$=17 K were observed in one sample, whereas reducing the sulfur content, shifted $T_C$ to 42 K. Both a-CS powders contain FM and PM extra magnetic phases which obscure their SC features. (iii) Magnetite and the PM impurities are absent in the a-C thin films grown by EBID. The SC transitions at 40 K [3] and at 34.4 K are sharp and more significantly, the Meissner effect is clearly observed in the FC branch. This may possibly stem from W which is embedded in the a-CS matrix. The role of the W inclusions (which do have a pairing potential) is to enhance the phase coherences by Josephson coupling nearby localized SC regions. The W islands do not contribute by themselves to the onset of SC in our films, since it appeared only after the sulfur treatment.

The appearance of the SC phases in a-CS reported here should be taken with a grain of salt because of the following reasons. (i) The low SC volume fraction obtained indicating that SC is not a bulk property, (ii) Not all the SC examples show the Meissner effect, which is probably caused by the PM impurity phases which cannot be avoided. (iii) That SC was not confirmed by transport measurements. (iv) The exact compositions of the various SC phases are not yet known. (v) The mechanism leading to these SC states is not yet known. However, the accumulate results presented here, prove unambiguously the existence of high $T_C$ SC phases in sulfur doped a-C materials, taken from different sources. Due to the tiny volume fraction observed in all samples any attempt to classify the nature the SC nature (s-type or p-type etc.) is misleading. It may be suggested that the nun-bulk SC traces in a-CS are due to long-sought-after interfacial mechanism proposed by Ginzburg in 1964 [54].This discovery shows that a-C materials are promising candidates for searching for new high $T_C$ superconductors.

Our results are in good agreement with other experiments on related graphite-based systems, and corresponding theoretical works, implying that structural disorder, topological defects, sulfur atom adsorption or phosphorous ion implantation, may trigger superconducting instabilities in graphite [1,6,30]. In particular, the experiments corroborate theoretical predictions, suggesting that both adsorbed sulfur and structural disorder can locally induce extra carriers (doping effect) into graphite/graphene and therefore trigger or enhance SC [30]. It should be pointed out, however, that in all these experiments the SC phase was localized (not percolating), the SF obtained were very low and the $T_c$ values significantly vary between the different samples.

*Peculiar magnetic behavior.* Three unusual reproducible magnetic features are observed a-CS samples regardless of whether the initial a-C powder was commercial or fabricated. (a) In contrast to the general trend where always FC > ZFC, a pronounced peak appears in the ZFC plots, and the FC curves cross the ZFC one (ZFC > FC). This observation is unique and to the best of our knowledge was not observed in the past (puzzle 1). (b) Moreover, unexpectedly the extra magnetization gained as a peak in the first ZFC sweep is washed out and disappears in the second ZFC run, performed a few minutes later (puzzle 2). (c) The pronounced peak in the ZFC branch is dynamic. The suppressed peak reappears after 18 months. This complexity does not permit measuring twice the same material or to compare results obtained on two aliquots although taken from the same batch. The speculations that experimental failures and/or adsorbed oxygen and Fe impurities caused these peculiar magnetic phenomena have been ruled out.

The rare behavior of ZFC>FC was observed in several ferri-magnetic systems such as $Sr_{0.8}La_{0.2}Ti_{0.9}Co_{0.1}O_3$ [23], $HoMnO_3$ [33], NiMnPd thin films [34] and in microporous carbon [1]. $Sr_{0.8}La_{0.2}Ti_{0.9}Co_{0.1}O_3$ undergoes a PM-ferri-magnetic phase transition at 50 K and the higher ZFC curve is a result of the superposition of PM and FM contributions to the total magnetic moment. In $HoMnO_3$ this phenomenon is also related to the ferri-magnetic structure of the two Ho and Mn sub-lattices. In NiMnPd thin films, it is attributed to antiferromagnetic (AFM) impurities which may be coupled to their FM host matrix and force the entire system to break into AFM domains coupled to each other, a model proposed by Imri and Ma [35]. ZFC > FC observed in microporous carbon at high external field (10 kOe), is attributed to the presence of meta-stable magnetic phases as proposed by Ro et al. [36]. None of these cases are applicable to a-CS.

The salient features shown for pyrolytic sulfur doped a-C samples are also observed in sulfur and boron doped commercial a-C materials. That means that these puzzles and unusual magnetic properties are not accidental, but rather common intrinsic properties of these a-C materials. The current state of experiments does not allow any consistent explanation to these magnetic phenomena. It is quite obvious that the two puzzles are connected to each other. Since the peaks are not reproducible and washed out after ZFC(i), they does not show up in the FC runs either. Therefore, due to the impurity phases, they rather lie above the ZFC(ii) curves and show their normal trend FC>ZFC(ii) as expected. The main questions remain: (a) what is the origin of the peaks and (2) why they appear in the first ZFC runs only.

We are not aware of any existing theoretical models that would be able to explain the peaks appearance. However, the experimental evidences accumulated so far indicate that structural disorder, the curvature of the carbon sheets, defects, or adsorbed foreign atoms (such as sulfur or boron) can be responsible for a local magnetic state of the $sp^2$-bonded carbon atoms in the graphitic structures. The presence of carbon rings other than six, may induce localization of the π-electron clouds and allow localization of unpaired spins, all can be behind the anomalous magnetic peak. Alternatively, it is reasonable to attribute these peaks to a tiny migration of sulfur to the sample surface caused by the external magnetic field, or to the presence of magnetite which in some way is coupled to the amorphous carbon atoms.

As a hand waving qualitative model, we may speculate that the a-CS materials are in the so called -two-state system- separated by a certain energy barrier which can be described by a double-well potential with a finite probability of finding the system in one of the two wells. Since a-CS is an inhomogeneous doped graphene material, in which the sulfur is distributed in a non-uniform manner, various magnetically ordered states (AFM, FM, spin-glass, RVB spin-singlet correlations), and even superconductivity can emerge. [37-38]. In particular, in the first ZFC process the system can be trapped in a one potential well magnetic state, whereas in the second ZFC process as well as in the FC regime, the system remains in the lower magnetization state. Whatever the explanation is, our exciting discovery of irreversibility in the first ZFC process is challenging for the theory of magnetism and propitious to further theoretical and experimental investigations.

In conclusion. It is shown above that the $T_C$ values of the various SC phase are very close to the elusive peak positions observed in the first ZFC branches. The occurrence of magnetic ordered states near the SC phases suggest that magnetism has a crucial role in the formation of the SC states. In that sense, it is possible that the a-CS materials are very similar to all other high $T_C$ systems in which SC emerges from magnetic states. It is hoped that, although showing only signatures of localized SC, this work would stimulate future studies to enhance $T_C$ and the volume fraction to the extent of achieving bulk high-temperature SC in such materials. Thus, systematic experimental work, aiming to increase the reproducibility as well as the SC volume fraction in sulfur doped a-C, is urgently needed.

**Acknowledgments**: The research is supported in part by the Israel Science Foundation (ISF, Bikura 459/09) and by the Klachky Foundation for Superconductivity. I thank Dr. E. Prilutskiy for providing the pyrolytic carbon materials, Prof. O. Millo for providing the a-C-W thin films and to Prof. Y. Kopelevich for his assistance in the early stage of this research.

**References**
[1] Y. Kopelevich, P. Esquinazi, J. H. S. Torres, and S. Moehlecke, J. Low Temp. Phys. **119**, 691 (2000).
[2] I. Felner and Y. Kopelevich. Phys. Rev. B. **79**, 233409 (2009).
[3] I. Felner, O. Wolf and O. Millo, J. Supercond. Nov. Magn. **25**, 7 (2012).
[4] I. Felner and E. Prilutskiy, J. Supercond. Nov. Magn: **25**, 2547(2012).
[5] I. Felner, O. Wolf and O. Millo, J. Supercond. Nov. Magn. **26**,514 (2013).

[6] A.M.Black-Schaffer and S. Doniach S, Phys. Rev. B **75**, 134512 (2007).
[7] R. Caudillo, X.Gao, R. Escudero, M. Jose-Yacaman, and J.B. Goodenough, Phys. Rev. B **74**, 214418 (2006).
[8] Li Zeng, E. Helgren, F. Hellman, R. Islam, David J. Smith and J. W. Ager III, Phys. Rev. *B* **75**, 235450 (2007).
[9] P. Esquinazi, D. Spemann, R. Höhne, A. A. Setzer, K.-H. Han, and T. Butz, Phys. Rev. Lett. **91**, 227201(2003).
[10] Y. Kopelevich, R. R. da Silva, J. H. S. Torres and A. Penicaud, and T. Kyotani, Phys. Rev. B **68**, 092408 (2003).
[11] P. Esquinazi, , R. Hohne, K.H.Han A. Setzer, D. Spemann and T. Butz, Carbon **42**, 1213 (2004).
[12] S. Bandow, T. Yamaguchi, and S. Iijima, Chem. Phys. Lett. **401**, 380 (2005).
[13] K. Tanaka, M. Kobashi, H. Sanekata, A. Takata, and T. Yamabe, S. Mizogami, K. Kawabata and J. Yamauchi, J. Appl. Phys. 71, 836 (1992).
[14]. K. Murata, and H. Ushijima, J. Appl. Phys. **79**, 978 (1996).
[15] A.V. Rode, E.G. Gamaly, A.C. Christy, J.F. Fitz Gerals, S.T. Hyde, R.G. Elliman, B. Luther-Davies, A.I. Veinger, J. Androulakis, and J. Giapintzakis, Phys. Rev. B. **70**, 054407(2004).
[16] R. Blinc, P. Ceve, D.Arcon, B.Zalar, A. Zorko, T. Apih, F. Milia, N. R. Madsen, A.G. Christy, and A.V. Rode, Phys. Stat. Sol. **13,** 3069 (2006).
[17] D. Arcon, Z. Jaglicic, A. Zorko, A. V. Rode, A.G. Christy, N.R. Madsen, E.G. Gamaly, and B. Luther-Davies, Phys. Rev. B. **74**, 014438 (2006).
[18] C. Mathioudakis and P. C. Kelires, Phys. Rev. B. **87**, 195408 (2013).
[19] A.I. Shames , A.M. Panich, E. Mogilko , J. Grinblat , E.W. Prilutskiy , E.A. Katz, Diamond and Related Mater. 16, 2039 (2007).
[20] H. Steinberg, Y. Lilach, A. Salant, O. Wolf, A. Faust, O. Millo and U. Banin, Nano Letters 9, 3671 (2009).
[21] J. Barzola-Quiquia and P. Esquinazi, J. Supercond. Nov. Magn. 23, 451–455(2010).
[22] R. R. da Silva, H. Torres and Y. Kopelevich, Phys. Rev. Lett. **87**, 147001 (2001).
[23] Yang Hai-Peng,Wen Hai-Hu, Zhao Zhi-Wen and Li Shi-Liang, Chin. Phys. Lett.
18, 1648 (2001).
[24] S. Moehlecke, Y. Kopelevich and M. B. Maple, Phys. Rev. B **69**, 134519 (2004).
[25] S. Bandow, T. Yamaguchi, and S. Iijima, Chem. Phys. Lett. **401**, 380 (2005).
[26] A.W. Mombru, H. Pardo, R. Faccio, O.F. de Lima, E.R. Leite, G. Zanelatto, A.J.C. Lanfredi, C.A. Cardoso, and F.M. Araujo-Moreira, Phys. Rev. B. **71**, 100404 (2005).
[27] O. Yuli, I. Asulin, G. Koren, O. Millo, and D. Orgad, Phys. Rev. Lett. **101**, 057006 ( 2008).
[28] W. Li, J. C. Fenton, Y. Wang, D.W. McComb, and P. A. Warburton, J. Appl. Phys. **104**, 093913 (2008).
[29] E. Majkova, S. Luby, M. Jergel, H. v. Lohneysen, C. Strunk, and B. George, Phys. Stat. Sol. (a) **145,** 509 (1994).
[30] G. Larkin and Y. Vlasov, Supercon. Sci. Tech. **24**, 092001(2011).
[31] C.W Chu, Physica C **482**, 33 (2012).
[32] V.L. Ginzburg, JETP **47,** 2318 (1964).
[33] E. Glastyan, B. Lorenz, K.S. Martirosyan, F. Yen, Y.Y. Sun, M.M. Gospodinov and C.W. Cue, J. Phys: Condens Matter **30**, 325241 (2008).
[33] Y. Oner, C.S. Lue, J.H. Ross, Jr., K.D.D. Rathayka, and D.G. Naugle, J. Appl. Phys. **89**, 7044 (2001).
[34] J. Imri, and S-K Ma, Phys. Rev. Lett **21**, 1399 (1975).
[35] C. Ro, G.S. Grest, C.M. Soukoulis, and K. Levin, Phys. Rev. B. **31**, 1682 (1985).
[36] A. J. Leggett, S. Chakravarty, A.T. Dorsey, M.P.A Fisher, A. Grag and W. Zwege, Rev. Mod. Phys. **59**, 1 (1987).
[37] Z. Y. Meng,T.C. Lang, S. Wessel, F.F. Assaad and A. Muramatso, Nature **464**, 847 (2010).